\documentclass[pra,aps,amsmath,twocolumn,showpacs,floatfix]{revtex4-1}
\usepackage{graphicx}
\usepackage{bm}
\usepackage{pdfpages}
\input{epsf}
\begin{document}
\epsfverbosetrue
\def\la{{\langle}}
\def\ra{{\rangle}}
\def\vep{{\varepsilon}}
\newcommand{\beq}{\begin{equation}}
\newcommand{\eeq}{\end{equation}}
\newcommand{\beqa}{\begin{eqnarray}}
\newcommand{\eeqa}{\end{eqnarray}}
\newcommand{\q}{\quad}
\newcommand{\e}{\ref}
\newcommand{\p}{\partial}
\newcommand{\A}{|\Omega'|}
\newcommand{\AC}{{\it AC}}
\newcommand{\n}{\\ \nonumber}
\newcommand{\om}{E}
\newcommand{\thet}{T}
\newcommand{\Om}{\Omega}
\newcommand{\tild}{\overline}
\newcommand{\Ee}{\mathcal{E}}
\newcommand{\os}[1]{#1_{\hbox{\scriptsize {osc}}}}
\newcommand{\cn}[1]{#1_{\hbox{\scriptsize{con}}}}
\newcommand{\sy}[1]{#1_{\hbox{\scriptsize{sys}}}}
%\draft

%\title{Loss and recapture of particles by a time dependent trap  near a continuum threshold }
\title{Adiabaticity in a time-dependent trap: a universal limit for the loss by touching the continuum }
\author {D. Sokolovski$^{a,b}$}
\author {M. Pons$^c$}
%\author {J. G. Muga$^{a,d}$}
\affiliation{$^a$ Departamento de Qu\'imica-F\'isica, Universidad del Pa\' is Vasco, UPV/EHU, Leioa, Spain}
\affiliation{$^b$ IKERBASQUE, Basque Foundation for Science, E-48011 Bilbao, Spain}
\affiliation{$^c$ Departmento de F\' isica Aplicada I, Universidad del Pa\' is Vasco, UPV-EHU, Bilbao, Spain}
%\affiliation{$^d$ Department of Physics, Shanghai University, 200444 Shanghai, China}
\date{\today}

\begin{abstract}
We consider a time dependent trap externally manipulated in such a way that one of its bound states is brought into an instant contact 
with the continuum threshold, and then down again. It is shown that, in the limit of slow evolution, the probability to remain in the bound state, $P^{stay}$ tends to a universal limit, and is determined only by the manner in which the adiabatic bound state approaches and leaves the threshold. The task of evaluating the $P^{stay}$ in the adiabatic limit can be reduced to studying the loss from a zero range well, and is performed numerically. Various types of trapping potentials are considered. 
Applications of the theory to cold atoms in traps, and to propagation of traversal modes in tapered wave guides are proposed.

\end{abstract}
\date{\today}
\maketitle
\vskip0.5cm
            % pls. do not remove this line

\section{Introduction}
Recent progress in laser-based techniques has led to the creation of various methods for trapping cold atoms. The laser induced potentials, used for this purpose, range from extended optical lattices \cite{Latt} to individual quasi-one dimensional traps \cite{Reiz}.
Such single traps, designed specifically in order to achieve single-site control, can be used, for example, for production of  atomic Fock states
\cite{Reiz1}. 
These states, containing a known number of atoms,  find numerous applications  in fields as diverse as metrology, few-body quantum physics \cite{Reiz3},\cite{Few1}, quantum entanglement \cite{Ent}, and quantum computing \cite{Reiz1}. Production of Fock states may be achieved by external manipulating of the trapping potential \cite{Reiz1}, \cite{Gon}-\cite{Fock4}, and the question of whether the adiabatic limit  is reached in  its evolution plays here a central role. The presence of continuum states makes the problem somewhat more complicated \cite{Tolst4}, \cite{US},\cite{US1} than the Landau-Zener case \cite{LAND}, where only two discrete levels are involved.
\newline
In a process similar to "laser culling" \cite{Reiz1} or "laser squeezing", \cite{Gon} the depth or the width of the trap is manipulated in such a way, that its bound states move closer to the continuum threshold, and some of the trapped particles are ejected into the continuum.
For example, in culling, if the trap is made shallower and then deeper again, an adiabatic bound state may make a U-turn before reaching the threshold.
If the evolution is slow, the adiabatic theorem (see, for example \cite{Adiab})  guarantees that the particle will remain trapped, and the probability of loosing it to the continuum will be exponentially small \cite{Tolst4}. 
Alternatively, the trap can be made so shallow that is temporarily ceases to support the original bound state, and only  recovers it once its depth begins to increase again. A slow evolution of this type will almost certainly lose the particle, as almost nothing will be recaptured by the deepening well. 
Separating the two regimes is the borderline case where the adiabatic bound state only touches the continuum threshold, and immediately resumes its downward journey. Relatively little is known about the probability to retain the particle within the trap if the threshold is approached slowly, and the "adiabatic" limit achieved in such evolution is the main subject of this paper.
\newline
Almost fifty years ago, Devadriani \cite{Devd} considered the chance to retain the particle in a three-dimensional zero-range (ZR) well whose magnitude, quadratic in time, vanishes at $t=0$. It was shown that the retention probability in this case is about $38\%$, and is independent of the speed of the evolution. More recently, it was demonstrated that this "$38\%$ rule holds in the 
adiabatic limit for any bound state of an arbitrary one-dimensional trap, subject to a similar quadratic-in-time evolution \cite{US1}.  
This result may suggest the existence of a universal adiabatic limit for the loss of particles 
caused by touching the continuum, at least 
in a one dimensional culling-like process. 
It would be reasonable to expect such a limit  to depend only of the manner in which an adiabatic bound state approaches and touches the threshold, 
and be common to all shapes of trapping potentials, masses of the particle, and to the ground and excited states alike \cite{US1}. 
In this paper we will demonstrate the existence of the limit, and evaluate the retention probabilities $P^{stay}$ for evolutions of different kinds. Our task is somewhat simpler than the one usually performed when the adiabatic limit is known beforehand, and the deviations from it are of interest. In what follows the subject is the limit itself, rather than the manner in which it may be reached.
%\newline

The rest of the paper is organised as follows. In Sect. II we will formulate the problem in the case of "culling". In Section III, in the adiabatic limit, we will reduce it to solving a time-dependent problem for a ZR potential, and demonstrate the existence of the universal limit for the bound states. Section IV contains the analysis of the resulting ZR problem.
 Sections V and VI contain brief reviews of existing analytical approaches to the problem. In Section VII we extend our analysis to evolutions of arbitrary type. In Sect. VIII we finally resort to numerical evaluation of $P^{stay}$, and check the validity of our conclusions for several realistic potentials. The case of "squeezing" is discussed in Sect. IX, and
Section X contains the summary of our results.
\begin{figure}
	\centering
		\includegraphics[width=5cm,height=6cm]{{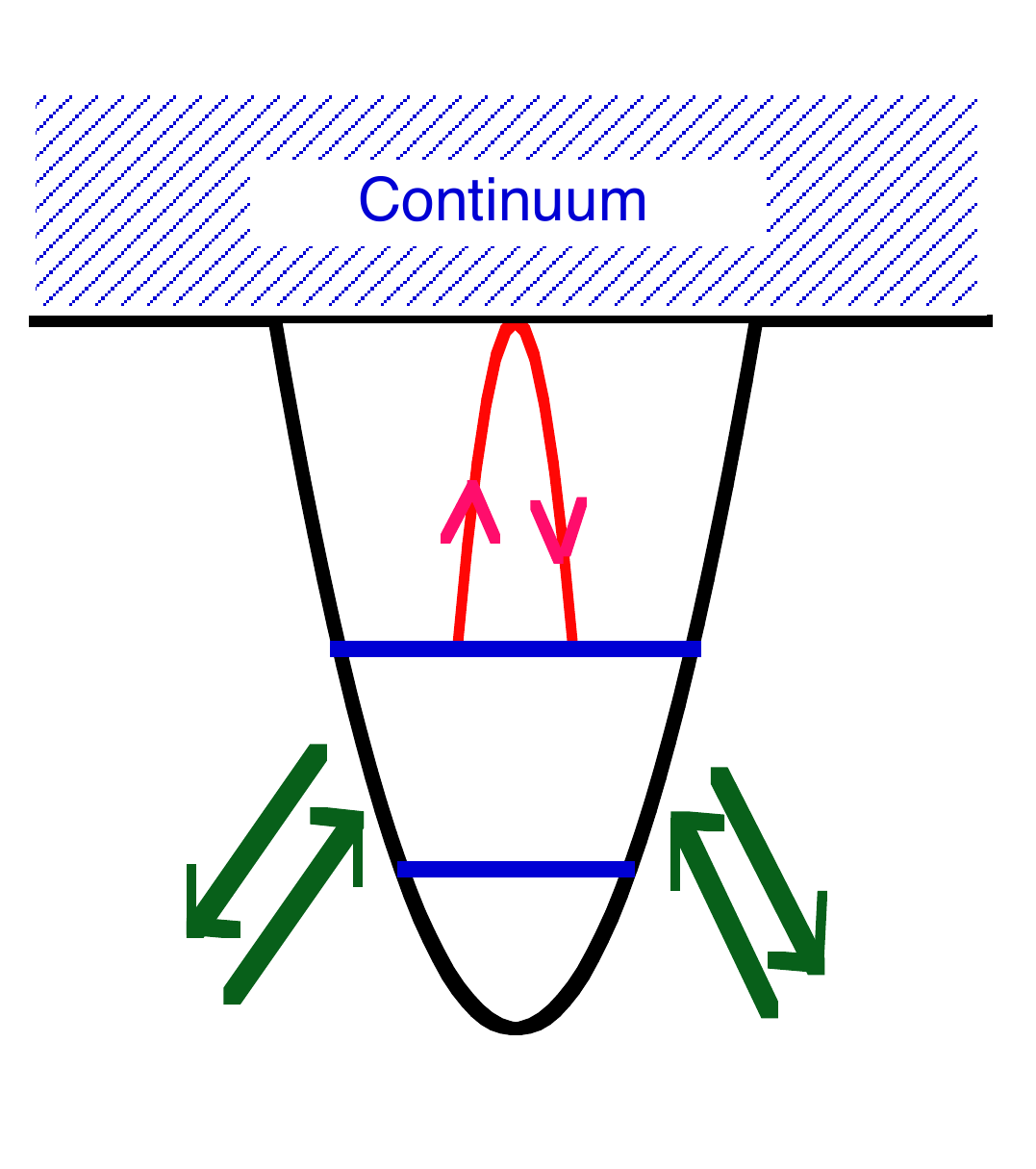}}
\caption{(Color online) Evolution of the potential $W(x,t)$ lifts its $n$-th adiabatic bound state, makes it briefly touch the continuum, 
and then brings it down again. We wish to evaluate the probability for a particle to remain in the state, $P_n^{stay}$, in the case the evolution is slow. }
\label{fig:0}
\end{figure}
\section {Loss by touching the continuum: "culling"}
We start by considering  the Schroedinger equation (SE) for a particle of a mass $\mu$ in a one-dimensional time dependent potential well  ($\hbar=1$), 
\begin{eqnarray}\label{a1}
i\p_t \Psi(x,t)=- \p^2_x \Psi/2\mu-W(x,t) \Psi, 
\end{eqnarray}
where 
\begin{eqnarray}\label{a2}
W(x,t)=(\rho^{(n)} +v^\nu |t|^\nu)W(x). 
\end{eqnarray}
The well has a finite range, so that $W(x)$, normalised by the condition $\int_{-\infty}^\infty W(x)dx=1$, vanishes for $|x|>a$. It may support several adiabatic states, $\phi_n(x,t)$, $n=0,1,2...$, with the energies $E_n(t)$, and the constant $\rho^{(n)}$ is chosen in such a way that $E_n(t=0)=0$. Thus, the time evolution of the potential brings the $n$-th adiabatic state up to the continuum threshold, and then brings it down the same way it came up.
The type of the evolution depends on the exponent $\nu$, which can be any positive real number. The speed of the evolution is controlled by the parameter $v$.
\newline
 In the spirit of the adiabatic theorem \cite{Adiab}, we wish to know how many particles, if any, will be lost to the continuum if the state is approaching the continuum very slowly, i.e., in the limit $v\to 0$. With this in mind, we will prepare the particle in the deep-lying $n$-th bound state at some large negative $t=-T$, 
 \begin{eqnarray}\label{a1a}
 \Psi(x,-T)=\phi_n(x,-T), \q T\to \infty,
 \end{eqnarray}
 and then evaluate the retention probability  $P_n^{stay}(v,\mu,\nu,W)$ to still find it in the same state at $t=T$.
 The probability is given by the square of the modulus of the corresponding amplitude, 
  \begin{eqnarray}\label{aa1}
P_n^{stay}(v,\mu,\nu,W)=\q\q\q\q\q\q\q\q\q\q\q\q\n
|\la \phi_n(T)|\exp[-i\int_{-T}^T\hat{H}(t)dt]|\phi_n(-T)\ra|^2, 
 \end{eqnarray}
where $\hat{H}$ is the operator in the r.h.s of Eq.(\ref{a1}), and the exponential is understood to be the time ordered product of 
non-commuting terms $\exp[-i\hat{H}(t)dt]$.  
Instead of evolving the initial state until $t=T$, we may evolve it to $t=0$, thus obtaining 
$\Psi(x,0)$. We will also need to evolve
the final state backwards in time to the same $t=0$. 
%Instead
By time reversal \cite{BAZBOOK}, this backward evolution can be replaced 
by a forward one, accompanied by the appropriate complex conjugation. Since $\phi_n(x,-T)=\phi_n(x,T)$, $\hat{H}(t)=\hat{H}(-t)$, and 
$\phi_n(x,T)=\phi^*_n(x,T)$, the result is $\Psi^*(x,0)$, and we have
  \begin{eqnarray}\label{aa2}
P_n^{stay}(v,\mu,\nu,W)=|\int \Psi(x,0)^2 dx|^2, 
 \end{eqnarray}
 which gives the probability $P_n^{stay}$ in terms of the wave function at $t=0$.
 %Thus, to evaluate the retention probability $P_n^{stay}$ it is sufficient to know the wave function at the moment the state touches the continuum. 
%We will retu (\e{aa2}), which helps reduce the computational effort, will be used in what follows.
%, making sure that $P^{stay}$ is indepen

%%%%%%%%%%%%%%%%%%%%%%%%%%%%%%%%%%%%%%%%%%%%%%%%%%%%%%%%%%%%%%%%%
\section {Reduction to the zero range model in the adiabatic  limit}
A scaling transformation \cite{FOOT} $t\to \alpha t$, $x\to \beta x$,  where 
\begin{eqnarray}\label{b4}
\alpha(\mu',v'|\mu,v)=(\mu'/\mu)^{1/(2\nu+1)} (v'/v)^{2\nu/(2\nu+1)},\q \q\n
 \beta(\mu',v'|\mu,v)=(\mu'/\mu)^{(\nu+1)/(2\nu+1)} (v'/v)^{\nu/(2\nu+1)},\q
\end{eqnarray}
converts Eq.(\ref{a1}) conditioned by (\ref{a1a}) into
\begin{eqnarray}\label{a3}
i\p_{t} \Psi(x,t)=- \p^2_{x} \Psi/2\mu'-(\frac{\alpha}{\beta}\rho^{(n)} +v'^\nu|t|^\nu)\tilde{W}(x') \Psi,
\end{eqnarray}
where $\tilde{W}(x')=\beta W(\beta x')$,
with a new initial condition
 \begin{eqnarray}\label{b5}
 \Psi(x,-T)=\beta^{-1/2}\phi_n(\beta x,-T), \q T\to \infty.
 \end{eqnarray}
Choosing $\mu'(v)=\mu(v'/v)^{1+\epsilon}$, with  $\epsilon > 0$, ensures that  in the adiabatic limit $v\to 0$  we have
% \begin{eqnarray}\label{b5}
$\beta=v^{-[1+\epsilon (\nu+1)/(2\nu+1)]}\to \infty$, 
while the constant term multiplying $W(x)$ vanishes, 
 %\begin{eqnarray}\label{b7}
$\alpha/\beta = (v/v')^{\epsilon \nu/2\nu+1}\to 0$.
 %\end{eqnarray}
As a result of the scaling, the well becomes narrower and deeper, while 
the initial distribution of the particle's positions also narrows, 
\begin{eqnarray}\label{b8}
\lim_{v\to 0}\tilde{W}( x')=\delta(x'), \q\q\q \q  \n\lim_{v\to 0}\lim_{T\to \infty}|\Psi(x,-T)|^2 \to
\beta^{-1}\delta(x'),
\end{eqnarray} 
where $\delta(x)$ is the Dirac delta.
\newline
Thus, after scaling,  we have to solve the SE for a zero-range (ZR) potential, 
\begin{eqnarray}\label{b9a}
i\p_{t'} \Psi(x',t')=- \p^2_{x'} \Psi/2\mu'-v'^{\nu}|t'|^\nu\delta(x')\Psi,
\end{eqnarray}
 which no longer depends on the particular shape of $W(x)$. 
 A scaling transformation cannot alter the value of a dimensionless quantity, so for the retention probability we should have
%This information is  lost also in the retention probability $P_0^{stay}$, for which we have
\begin{eqnarray}\label{b10}
P_0^{stay}(v\to 0,\mu,\nu,W)=P_\delta^{stay}(\nu),\q\q
\end{eqnarray} 
where $P_\delta^{stay}(\nu)$ is the retention probability for a very heavy particle,  $\mu'\to \infty$, trapped in a ZR well evolving at a rate $v'$.
Since $v'$ was chosen arbitrarily,  $P_\delta^{stay}(\nu)$, should not depend on its choice, if  we 
expect Eq.(\ref{b10}) to be correct. Next we will show that this is, indeed, the case. 
%%%%%%%%%%%%%%%%%%%%%%%%%%%%%%%%%%%%%%%%%%%%%%%%%%%
\section {The loss from a zero-range well}
It is a simple matter to check that the transformations 
%of the scaling group
(\ref{b4}) leave the form of the SE for a ZR well 
(\ref{b9a}) unchanged, except for replacing $\mu'\to \mu''$, and $v'\to v''$. 
For $\rho < 0$,  a ZR well $\rho(t) \delta(x)$ supports  a single adiabatic bound state with an energy $E_0(t)=-\mu \rho(t)^2/2=-\mu v^{2\nu}|t|^{2\nu}/2$ (see the inset in Fig.2)
\begin{eqnarray}\label{d1}
\varphi_0(x,t)=\sqrt{-ik_0(t)}\exp[ik_0(t)|x|],\n
 k_0(t,\mu,v)=-i\mu \rho(t)=i\mu v^\nu|t|^{\nu}.
\end{eqnarray}
%Thus, for a ZR well, t
Thus, the initial condition (\ref{a1a}) also remains unchanged under the scaling (\ref{b4}),
%\begin{figure}
%	\centering
%		\includegraphics[width=8cm,height=6cm]{{FIG1.pdf}}
%		\includegraphics[width=7.5cm,height=5cm]{{FIGSQRpot1.pdf}}
%\caption{(Color online) Adiabatic energy of the bound state of a zero range potential
%for different kinds of the evolution.}
%\label{fig:0}
%\end{figure}
acquiring only an inessential constant factor,
\begin{eqnarray}\label{d2}
\varphi_0(x,T)\to \beta^{-1/2}\varphi_0(x,T).
\end{eqnarray}
Since scaling leaves the values of dimensionless quantities unaltered, $P^{stay}_\delta $ may not depend on the choice of the particle's mass $\mu$, or the speed of evolution $v$, 
\begin{eqnarray}\label{d3}
P_\delta^{stay}(v,\mu,\nu)=P_\delta^{stay}(\nu), 
\end{eqnarray}
which is the desired result.
%(see the inset in Fig....)
\begin{figure}
	\centering
		\includegraphics[width=9cm,height=7cm]{{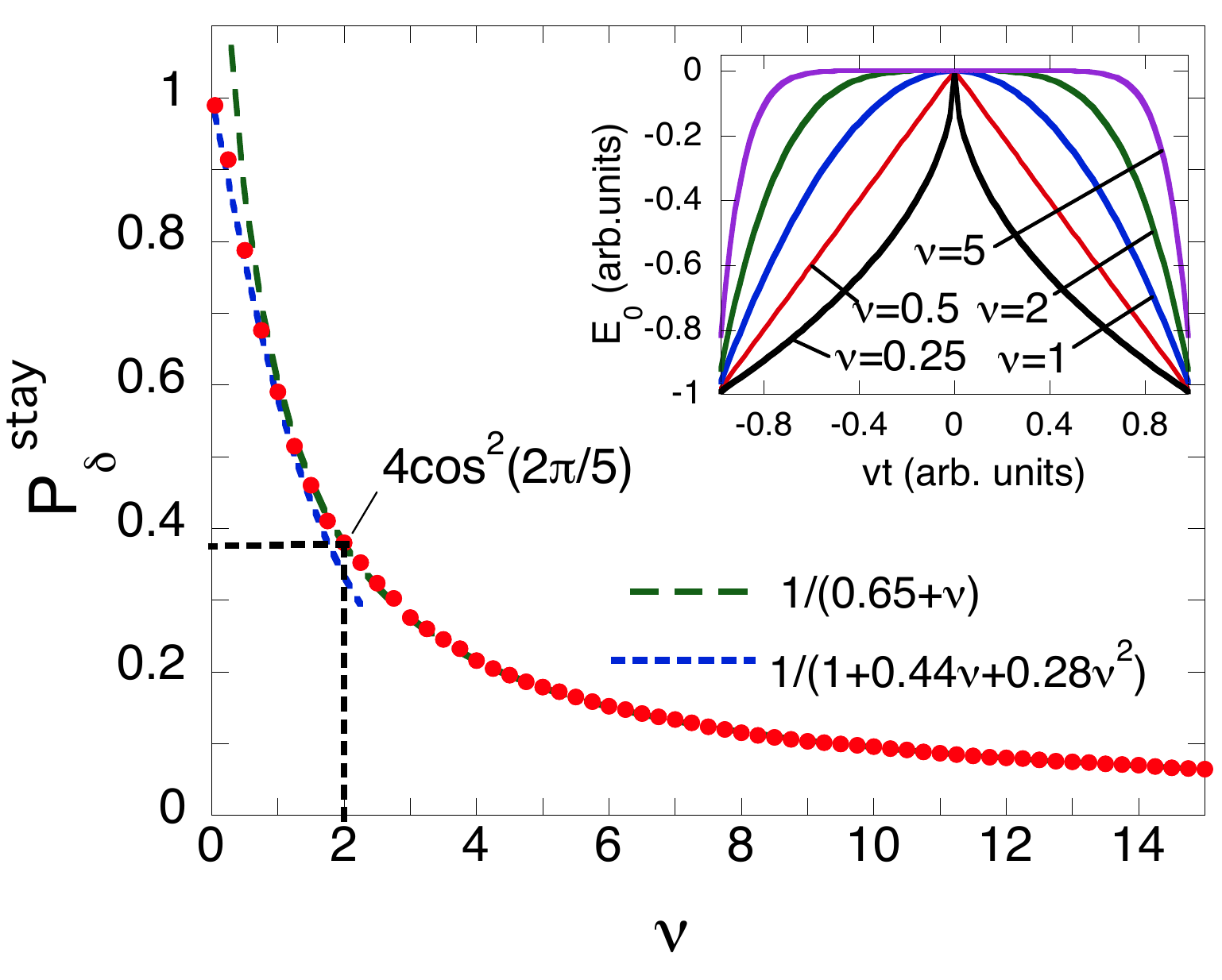}}
\caption{(Color online) Probability for a particle to remain in the bound state of a zero-range well $W(x,t)=-v^\nu|t|^{\nu}\delta(x)$, $P^{stay}_{\delta}(\nu)$, obtained by numerical integration of Eq.(\ref{b9a}). Also shown are the approximations (\ref{g0}) and  (\ref{g00}), as well as the value $P^{stay}_{\delta}(\nu=2)\approx 0.38$, which can be obtained analytically \cite{Devd}, \cite{US1}.
%The probability is independent of both $\mu$ and $v$, as illustrated in the inset for $\nu=1.75$.
The inset shows the adiabatic energy of the state 
for evolutions of different kinds.}
\label{fig:3}
\end{figure}
\newline
To provide an additional check,  we note also that probability densities, such as the energy distribution of the emitted particles, $w(E)$,
must change when transformations (\e{b4}) are applied. With $t$ scaling as $\alpha t$,
%and the product $Et$ unaltered, 
 the energy 
must scale as $E \to E/\alpha$,  and we should have (for a more detailed proof see also the Appendix A)
%Applying  $\mu' \to \mu''$ and $v'\to v''$, we should have (see the Appendix for details)
\begin{eqnarray}\label{d3a}
w(E|\mu'',v'')=\alpha(\mu'',v''|\mu',v')^{-1} 
\n
\times w(E/\alpha(\mu'',v''|\mu',v')|\mu',v'), 
\end{eqnarray}
which, since $P^{stay}_\delta=1-\int w(E)dE$, would confirm 
the validity of (\ref{d3}) (cf. Fig.3).
\begin{figure}
	\centering
		\includegraphics[width=8.5cm,height=6.5cm]{{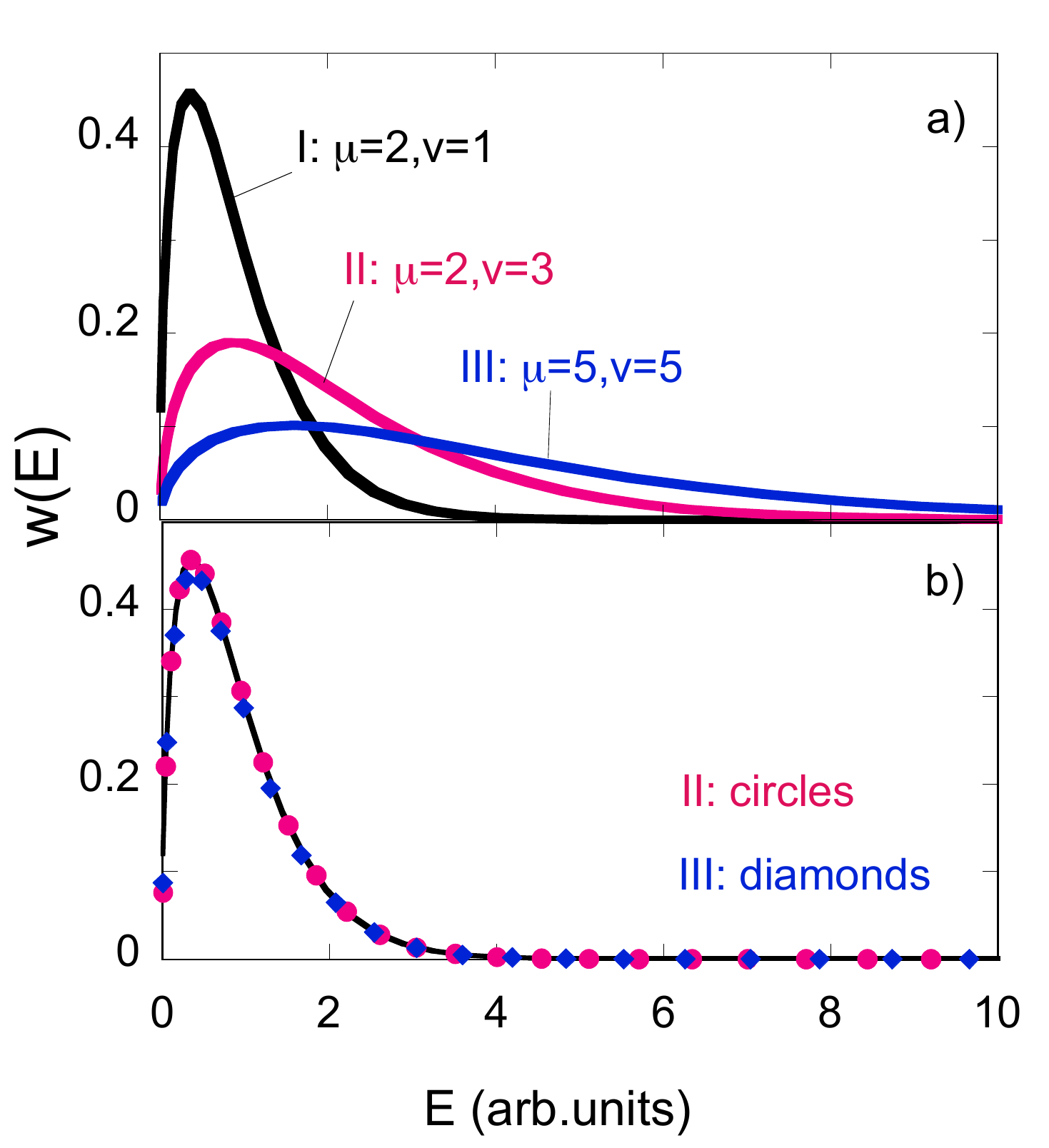}}
\caption{(Color online) a) Energy distribution $w(E)$ (arb. units) of the particles ejected from a zero-range well $ v^{2}|t|^{2}\delta(x)$ for 
various masses $\mu$ and speeds of evolution $v$, obtained by numerical integration of Eq.(\ref{b9a}); b) the result of scaling the curves $II$ and $III$ as prescribed by Eq.(\ref{d3a})  }
\label{fig:3}
\end{figure}
%\newline
%%%%%%%%%%%%%%%%%%%%%%%%%%%%%%%%%%%%%%%%%%%%%%%%%%%
\newline
The limit (\ref{b10}) can now be seen as universal in the following sense: in a "culling" process, the loss by touching the continuum in a slowly evolving well is determined only by the manner in which the state approaches the threshold, i.e., on the exponent $\nu$.
It is independent of the shape of the well and the particle's mass, and equals the loss from a ZR well. 
\newline
It remains to obtain the function $P^{stay}_\delta (\nu)$, preferably in the most general and transparent way. In the next two Sections we review two of the analytical approaches available for the problem, mostly to illustrate the difficulty one faces 
in doing so.
%%%%%%%%%%%%%%%%%%%%%%%%%%%%%%%%%%%%%%%%%%%%%%
\section {Zero range well: the Sturmian approach}
The main difficulty in solving equation (\e{a1}) analytically is the presence of the continuum states, also affected by the change of the potential well.
One way to simplify the problem is to use the discrete basis of the Sturmian states satisfying the outgoing waves boundary conditions \cite{US}, \cite{US1}, \cite{ST1}-\cite{ST4}. 
For a real energy $\om$, 
the method finds 
%a potential $\rho W(x)$, or more precisely, 
the values $\rho_n(E)$,
real or complex valued, such that the potential $\rho_n W(x)$ supports a suitable Sturmian state. 
The case of a ZR potential, $W(x)=\delta(x)$,  is particularly simple: there is only one one such value, $\rho_0=i\sqrt{2\om}$, and a single
Sturmian, so that the solution of Eq.(\e{a1})  with $v=\mu=1$ can be found in the form 
\begin{eqnarray}\label{c1}
\Psi(x,t) = \int d\om \exp(-i\om t) B(\om)S_0(x,\om), 
\end{eqnarray}
where $S(x,\om)$ is the Sturmian function, 
\begin{equation}\label{c2}
S_0(x,\om)=\exp(i\sqrt{2\om}|x|).
\end{equation}
In Eq.(\e{c1}) the integration contour runs just above the real axis on the first sheet of the Riemann surface of $\sqrt{\om}$ cut along the positive semi-axis, where, as a function of $x$, $S_0(x,\om)$ decays for $\om < 0$, and oscillates for $\om >0$ \cite{US1}. 
Inserting (\e{c1}) into (\e{b9a}) yields the equation for the unknown function $B(\om)$, 
\begin{equation}\label{c3}
\int \mathcal{G}(\om-\om') B(\om')d\om'+i\sqrt{2\om}B(\om)=0, 
\end{equation}
where 
%$\rho(\om)=i\sqrt{2\om}$ and 
the kernel $\mathcal{G}(\om)$ is formally defined as the Fourier transform of $|t|^\nu$
\begin{equation}\label{c4}
\mathcal{G}(\om)= (2\pi)^{-1}\int |t|^\nu \exp(i\om t) dt. 
\end{equation}
Equation (\e{c3}) is most useful whenever  $\nu$ is an integer. For an odd $\nu$, $\nu=1,3,5...$ we may replace $|t|$ with $-t$,
follow the evolution until $t=0$, obtain $\Psi(x,0)$, and then evaluate the integral (\e{aa2}).
With this, the kernel (\e{c4}) becomes $\mathcal{G}(\om)=-(-i)^\nu \partial^\nu_\om \delta(\om)$, and after integrating by parts 
Eq.(\e{c3}) yields
\begin{equation}\label{c5}
%(-1)^{\nu+1}
i^\nu \partial^\nu_\om B+i\sqrt{2\om}B=0. 
\end{equation}
In the simplest case of the evolution linear in time, $\nu=1$, the resulting first order equation is easily solved, 
yielding $\Psi(x, t\le0)$ given by the quadrature (\ref{c1}) \cite{US1}.
\newline
For an even $\nu$, $\nu=2,4,6,...$ we can replace $|t|$ with $t$, and again obtain for $B(\om)$  Eq.(\e{c5}).
%\begin{equation}\label{c6}
%i^\nu \partial^\nu_\om B(\om)+i\sqrt{2\om}B(\om)=0. 
%\end{equation}
In the quadratic-in-time case, $\nu=2$, Eq.(\e{c5}) is of the second order,  
%formally resembles the SE for a fictitious "particle" scattered 
%by a complex valued "potential" $\rho(\om)$ \cite{US}. It 
has an exact solution expressed in terms of the Hankel functions, and gives 
a retention probability (\e{aa1}) of about $38\%$ \cite{Devd}, \cite{US}. 
For a even integer $\nu>2$ one faces a similar but more difficult task of finding the correct boundary condition for the 
ordinary differential equation (ODE) (\ref{c5}), and expressing the solution in terms of "incoming" and "outgoing" waves for 
$E\to -\infty$ \cite{US1}. 
In the general case of a non-integer $\nu$, the kernel $\mathcal{G}(\om)$
is obviously related to a fractional derivative of the $\delta$-function (see, for example, \cite{FRACT}), which makes the resulting fractional order ODE satisfied by $B(\om)$ even less tractable,

%While reducing the partial differential equation (\e{a1}) to an ordinary one whenever the exponent $\nu$ is an integer, 
%the Sturmian approach offers little additional insight  into the problem in the general case. 
%%%%%%%%%%%%%%%%%%%%%%%%%%%%%%%%%%%%%%%%%%%%%%
\section {The Siegert-state approach}
A different more general approach, based on expansion of the time dependent state in terms of Siegert rather than Sturmian states has 
recently been developed in \cite{Tolst4}, \cite{Tolst0}-\cite{Tolst2}  for a class of finite range potentials which vanish for $|x|\ge a$. Next we will explore its usefulness for treating the ZR problem at hand. In our case, for a given (real) potential $\rho W(x)$ the method looks for the Siegert states associated with the poles $k_n(\rho)$
 %of the scattering matrix, in this case 
 of the transmission and reflection amplitudes, $T$ and $R$, in the complex plane of the momentum $k=\sqrt {2\mu E}$. The technique is based on imposing outgoing waves boundary condition at $x=\pm a$, and expanding the wave function $\Psi_I(x,t)$ in the inner region, $|x| < a$, in terms of the corresponding (Siegert) eigenstates. Once the wave function in the inner region is known, the solution in the outer regions $x> a$ and $x<-a$, $\Psi_O(x,t)$, is obtained by solving there the free-particle SE with the boundary conditions $\Psi_O(\pm a,t)=\Psi_I(\pm a,t)$. 
% In this way the method provides 
%the wave function $\Psi(x,t)$ on the whole $x$-axis, and avoids the difficulties arising from the divergent behaviour of the anti-bound and incoming Siegert states as $x\to \pm \infty$ \cite{Tolst}. 
Treatment of the wave vector $k$ rather than the energy $E=k^2/2\mu$ as an eigenvalue requires linearisation of the problem \cite{Tolst0}, doubling the dimension of the Hilbert space in the inner region, and introduction 
of the fractional time derivative, $\hat{\lambda}_t=\exp(3\pi i/4)\sqrt{2\partial_t}$.

The transmission amplitude for the ZR potential $\rho \delta(x)$, is well known \cite{REVDELT} to be $T(k,t)=k/(k+i\mu\rho)$. 
Associated with the pole at $k=-i\mu\rho$ is the single Siegert state (\ref{d1}). 
As the width of the well $a$ tends to zero, the inner region contracts to a single point, $x=0$, 
and the method allows us to find $\Psi_I(0,t)$, which is sought in the form [cf. Eq.(29) of \cite{Tolst4}]
\begin{eqnarray}\label{dd2}
\Psi_I(0,t)=a_0(t)\varphi_0(0,t)=\sqrt{-ik_0(t)}a_0(t).\n
\end{eqnarray}
The unknown function $a_0(t)$ satisfies Eqs.(30)-(31) of \cite{Tolst4} [without the factor of $2$ in denominator of Eq.(31), since 
our problem is on the whole $x$-axis \cite{Tolst2}]. In the limit $a\to 0$, and with $m=n=0$, they reduce to a single equation for $\Psi_I(0,t)$
\begin{eqnarray}\label{dd3}
\hat{\lambda}_t\Psi_I(0,t)-ik_0\Psi_I(0,t)=0.
\end{eqnarray}
 Taking the Fourier transform, $\Psi_I(0,t)=\int d\om \exp(-i\om t) \Psi_I(0,\om)$, and recalling that
 $\hat{\lambda}_t\exp(-i\om t)=i\sqrt{2\om}\exp(-i\om t)$ \cite{Tolst4}, for $\mu=v=1$ we have
\begin{equation}\label{dd4}
i\sqrt{2E}\Psi(0,\om)+\int \mathcal{G}(\om-\om') \Psi(0,\om')d\om'=0, 
\end{equation}
where $\mathcal{G}(\om)$ is defined by Eq.(\e{c4}). Comparison with Eqs. (\ref{c1}), (\ref{c2}) and (\ref{c3}) shows that we arrived at the equation for 
the value of the wave function at the origin $\Psi(x=0,E)$ obtained earlier in the Sturmian approach and, therefore, face the same problem of solving it. The result is not unexpected. There is a close relation between the Sturmian eigenvalues $\rho_n$ and the Siegert energies $E_n=k_n^2/2$, similar to the relation between the Regge and complex energy poles of a scattering matrix (see, e.g., \cite{PCCP}). While the Sturmian approach of Sect. VI uses the analytical function $\rho_0(E)=i\sqrt{2E}$, the Siegert-state method 
employs its inverse, $E_0(\rho)= -\rho^2/2$, and both techniques lead in the end to the same equation.
% (\ref{c3}) and (\ref{d4}).
%%%%%%%%%%%%%%%%%%%
\section{Universality of the  adiabatic limit}
Even though the Siegert-state approach does not offer an analytic solution to the problem, it allows us 
to prove the validity of  Eq.(\ref{aa2}) beyond the particular type of evolution considered so far. Next we will show that, 
in the adiabatic limit, the loss by touching the continuum always depends only on the manner in which 
an adiabatic eigenstate approaches the continuum threshold. For the adiabatic energy $E_n(t)$ at $t\approx 0$, without loss of generality, we write
\begin{eqnarray}\label{q1}
%lim_{v\to 0}
E_n(vt)= k_n(vt)^2/2\mu \approx  - Cv^{2\nu}|t|^{2\nu}, \q C>0=const.\q\q
% \q C>0=const , 
\end{eqnarray}
It is necessary to demonstrate that it is the power $\nu$ alone, which determines the loss to the continuum as $v\to 0$, and 
the proof is as follows. As $v\to 0$, we may  neglect all Siegert states \cite{Tolst4}, except the $\phi_n$, which is to touch the continuum,  
%Following \cite{Tols4} we 
and look for the solution in the form 
\begin{eqnarray}\label{q2}
\Psi(x,t)\approx a_n(t,\nu,v)\phi_n(x,t). 
\end{eqnarray}
The coefficient  $a(t,\nu,v)$ satisfies Eq.(30a) of Ref.\cite{Tolst4}, which in the limit $v\to 0$, reduces to [cf. Eqs.(31), (34)  and (36) of \cite{Tolst4}]
 \begin{eqnarray}\label{q3}
\hat{\lambda}_ta_n(t,v)-ik_n(vt,\nu)a_n(t,v)=0, 
\end{eqnarray}
with $k_n(vt)=i \sqrt{2\mu C} v^\nu |t|^{\nu}$. For $v\to 0$, the factor $\sqrt{\mu C}$, determines how rapidly $a(t)$ tends to its limit
$a_n(t,\nu,v\to 0)$, but not the limit itself. The limit must, therefore, be the same for a particle of any mass, and for any $C$ in Eq.(\ref{q1}). Finally, since $\phi_n(x,t)$ is normalised to unity, $\int \phi^2_n(x,t)=1$, [cf. Eq.(22) of \cite{Tolst2}], 
and insertion of (\ref{q2}) into Eq.(\ref{aa2}) yields
\begin{eqnarray}\label{q4}
P_n^{stay}(v\to0,\mu,\nu,W)\to  |a_n(t\to 0,\nu,v\to 0)|^4, 
\end{eqnarray}
Since we have shown that the r.h.s. of Eq.(\ref{q4}) depends only on the power $\nu$ in Eq.(\ref{q1}),
Eq.(\ref{b10}) must hold in general, for a particle of any  mass, and  for any state of a finite range well of any shape. 
Note that the argument can be extended to the case of an asymmetric evolution, where 
the state approaches  the continuum and then leaves it in a different manner (see Appendix B).
\newline
The above still does not offer a simple way for calculating  $P^{stay}(\nu)$ as $v\to 0$ analytically, but makes us free to choose the simple ZR model  for the purpose. The corresponding SE (\e{b9a}) can be easily solved numerically, 
and we will do it in the next Section.
%%%%%%%%%%%%%%%%%%%%%%%%%%%%%%%%%%%%%%%%%%%%%%%%%%%%%%%%%%%
\section{The universal adiabatic limit}
 Equation (\ref{a1}) is solved by the finite differences method \cite{NUM} for a particle of $\mu=1$ in a ZR well $W(x,t)=-v^{\nu}|t|^{\nu}\delta(x)$ placed between two infinite walls at $x=\pm L$. Since the solution is symmetric around the origin, it is sufficient to consider only the right half-space, with the boundary conditions $\partial_x\log(\Psi(0,t))=-v^{\nu}|t|^\nu$ and $\Psi(L)=0$ at $x=0$ and $x=L$, respectively. The initial condition (\ref{a1a}) is imposed at $T$ large enough to make
$P^{stay}_\delta(\nu)$ independent of the choice, and $L$ is chosen sufficiently large to avoid unphysical reflections.
The calculation is made easier by the freedom of choosing $v$ without changing the value of $P^{stay}_\delta$, 
which is then obtained with the help of Eq.(\ref{aa2}). The results are shown in Fig.2, which is the central result of this paper.
We note that for $\nu\gtrsim 1$ $P^{stay}_\delta(\nu)$ is reasonably well described by a rational function 
\begin{eqnarray}\label{g0}
P^{stay}_\delta(\nu) \approx [0.65+\nu]^{-1}
%\frac{1}{0.65+\nu}, 
\end{eqnarray}
while for $\nu \lesssim 1$ 
%we find
\begin{eqnarray}\label{g00}
P^{stay}_\delta(\nu) \approx 
[1+0.44\nu+0.28\nu^2]^{-1}
%\frac{1}{1+0.44\nu+0.28\nu^2}, 
\end{eqnarray}
provides a suitable approximation.
\newline
Validity of Eq.(\ref{b10}) is also checked numerically for the particles trapped in the ground and excited states of three potentials
($\theta(x)=1$ for $x\ge a$ and $0$ otherwise),
\begin{eqnarray}\label{g1}\n
W_I(x)=(2a)^{-1}\theta(x-a)\theta(x+a), \q\q \text {square well,}\q\q\n
W_{II}(x)=(4a^3/3)^{-1}(a^2-x^2)\theta(x-a)\theta(x+a), \text { parabolic, }\q\q \n
W_{III}(x)=(2a^2)^{-1}(a-x)\theta(x-a)\theta(x+a), \q \text {asymmetric,}\q\q
\end{eqnarray}
and the results are shown in Fig.4 for different values of $\nu$ and $\mu$.
\begin{figure}
	\centering
		\includegraphics[width=9cm,height=11cm]{{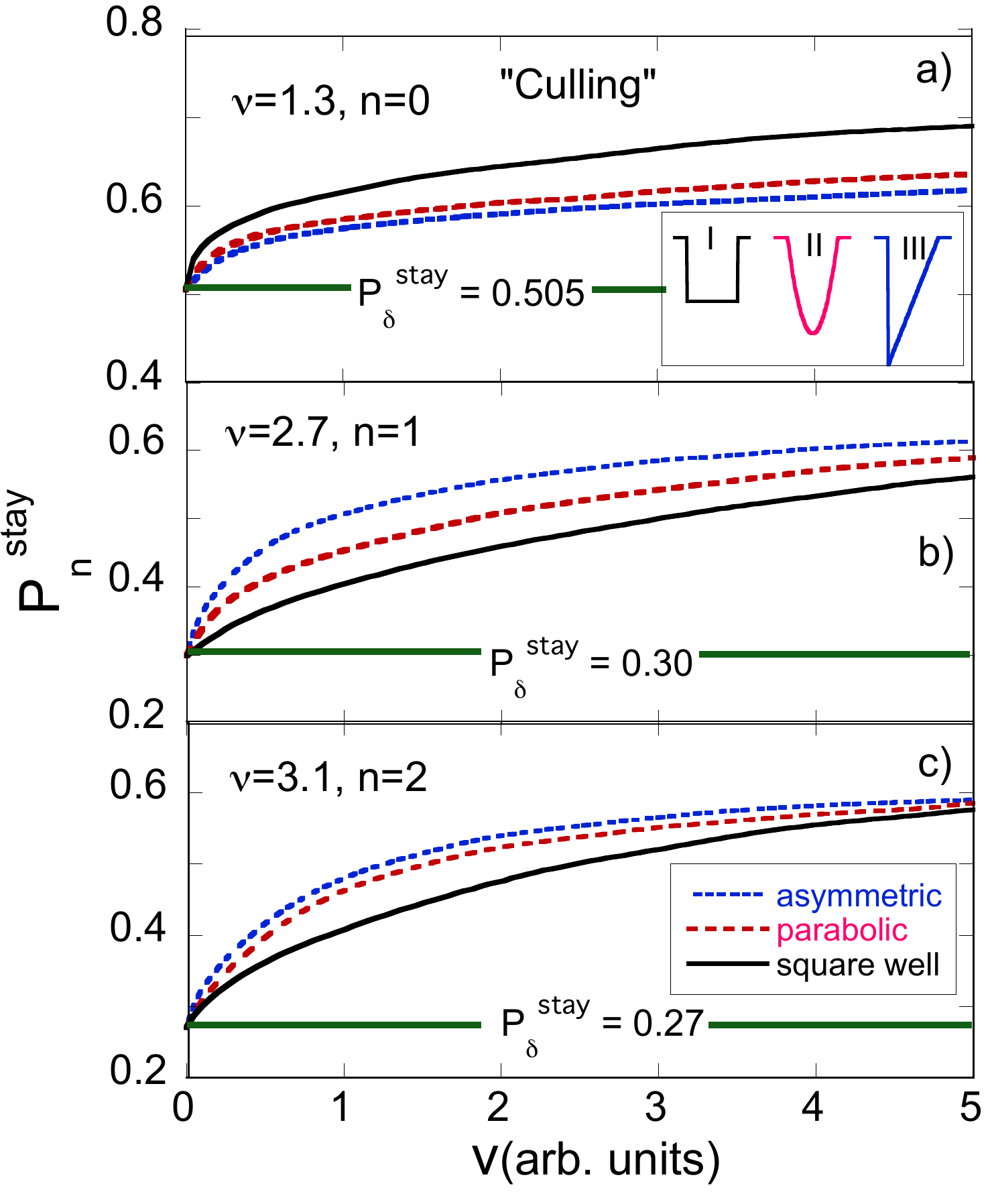}}
\caption{(Color online) Culling: the probability to remain in the $n$-th excited  state, $P_n^{stay}$ vs. $v$ (arb. units),  for the three potential wells
(\ref{g1}) shown in the inset: a) $\mu=1$, $\nu=1.3$, and $n=0$; b) $\mu=1$, $\nu=2.7$, and $n=1$; c) $\mu=1$, $\nu=3.1$, and $n=2$.
Also shown by horizontal lines are the corresponding results for the zero-range well.}
\label{fig:3}
\end{figure}
%%%%%%%%%%%%%%%%%%%%%%%%%%%%%%%%%%%%%%%%%%%%%%%%%%%%%%%%%%%
\section{Loss by touching the continuum: "squeezing"}
A different type of evolution, called squeezing in \cite{Gon}, proceeds by making the trap narrower, while leaving its depth unchanged. 
%One would ask whether the adiabatic limits obtained in the previous Section would also apply in the case of squeezing?
%We write the squeezed potential for the form
Before concluding, we will show that the chance to remain in a bound state brought to a brief contact with the continuum by a squeezed potential
\begin{eqnarray}\label{r1}
W(x,t)=W(x/L(t))=W(x/(L^{(n)}+v^\nu t^\nu)), \n
 E_n(t=0)=E_n(L^{(n)})=0,\q\q\q\q\q\q\q\q
\end{eqnarray}
is the same as in the case of culling (\ref{a1}). As discussed in Sect.VII, it is sufficient to demonstrate that in both cases a bound state approaches the continuum in the same manner. Let $E_n(\rho,L)$ be the energy of the bound state in a potential $\rho W(x/L)$. In the case of culling (\ref{a1}), we have $\rho(t)=\rho^{(n)}+v^{\nu} |t|^{\nu}$ and $L=const$,  so that 
\begin{eqnarray}\label{r3}
%lim_{v\to 0}
E_n(t) \approx -C(\rho-\rho^{(n)})^2 = - Cv^{2\nu} |t|^{2\nu}, \q\q
\end{eqnarray}
where $C= \partial^2_\rho E(\rho^{(n)},L)/2$.
 Consider the SE describing a bound state in a  potential $\rho W(x/L)$, written in some dimensionless variables, 
\begin{eqnarray}\label{r4}
[- \p^2_x/2-\rho W(x/L) -E_n(\rho, L)]  \phi_n(x,\rho, L)=0. 
\end{eqnarray}
By scaling the $x$-variable, $x \to xL'/L$, 
%an putting $\rho=1$ 
we can express $E_n$ for a squeezed well of a fixed depth $\rho$ in terms of that for a culled well of a fixed width $L'$,
\begin{eqnarray}\label{r5}
E_n(\rho,L)=\gamma^{-2}E_n(\gamma^2\rho, L'), \q \gamma\equiv L/L'.
\end{eqnarray}
Let $\rho^{(n)}(L')$ and $L^{(n)}(\rho)$ be the values of the corresponding parameters, for which the bound state disappears, 
$E_n(\rho^{(n)}, L')=0$, and $E_n(\rho, L^{(n)})=0$
It follows that $L^{(n)}(\rho)=L'\sqrt{\rho^{(n)}(L')/\rho}$, and expanding $E_n(\rho,L)$ in Eq.(\ref{r5}) around $L=L^{(n)}(\rho)$ yields
\begin{eqnarray}\label{r6}
E_n(\rho,L)\approx -C'[L-L^{(n)}(\rho)]^2=-C'v^{2\nu} |t|^{2\nu} ,\q
\end{eqnarray}
where
%\begin{eqnarray}\label{r7}
$C'=-\partial_L^2E_n(\rho,L^{(n)})/2$ $=-2\rho^2\partial^2_\rho E(\rho^{(n)},L')/L'^2$.
%\end{eqnarray}
Equations (\ref{r3}) and (\ref{r6}) differ only by inessential constant factors, and the loss by touching the continuum must be the same for the potentials in Eq. (\ref{a2}) and (\ref{r1}). This result is easily verified numerically, as Fig.5 shows.
\begin{figure}
	\centering
		\includegraphics[width=9cm,height=6cm]{{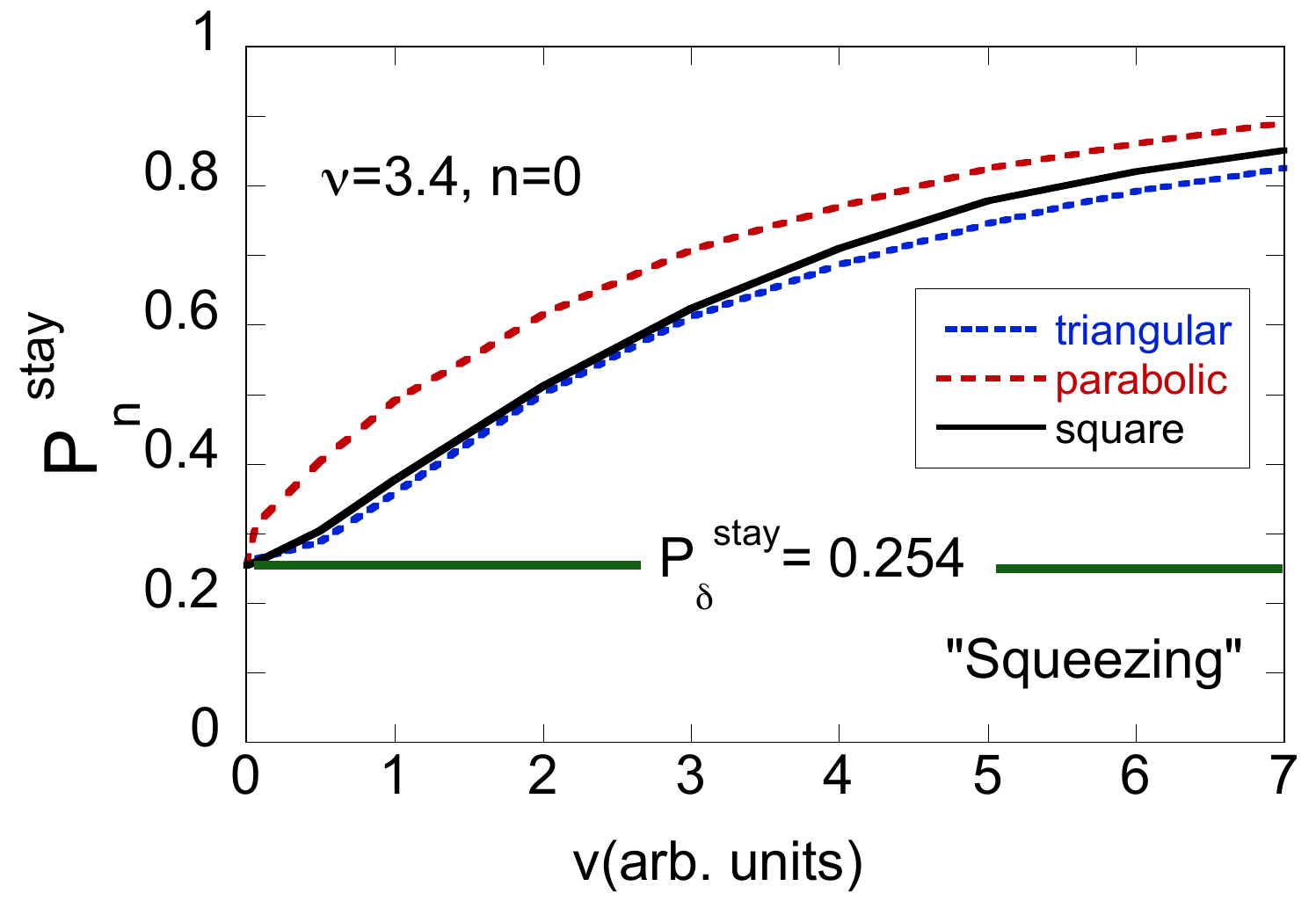}}
\caption{(Color online) Squeezing: the probability to remain in the ground  state, $P_0^{stay}$ vs. $v$ (arb. units),  for the three potentials
(\ref{g1}), with $\mu=1$, $\nu=3.4$, and $n=0$. Also shown by a horizontal line is the corresponding result for the zero-range well.}
\label{fig:3}
\end{figure}
%%%%%%%%%%%%%%%%%%%%%%%%%%%%%%%%%%%%%%%%
\section{Summary and discussion}
In one dimension,  there exists a universal adiabatic limit for the probability to remain in a bound state of a slowly evolving trap, $P^{stay}$, as the state briefly touches the continuum threshold. 
The limiting value of the $P^{stay}$ is determined only by the manner in which the state approached the continuum, and is independent of the particle's mass, the particular shape of the trapping potential, or the details of the trap's evolution.
More precisely, if the adiabatic energy of the particle near the threshold changes as $E_n(t) \approx -(v|t|)^{2\nu}$, then
for $v\to 0$ the probability $P^{stay}(\nu)$ tends to the adiabatic limit, which depends only on the exponent $\nu$. In a way, this an expected result.
If the evolution is slow, the particle is exchanged between the bound state, while it is close to the threshold $E=0$, and the few low-lying continuum states. The presence of other bound states in the well and the overall structure of the continuum should, therefore, play no role for the outcome of this exchange.  
%\newline

Evaluation of the limiting values of $P^{stay}(\nu)$ is a separate matter. Given that the result should hold for all potentials, we may take 
the simplest case of a zero range well as a reference. With the help of either Sturmian, or Siegert state approach, the problem can be reduced to solving an ordinary differential equation. However, in the general case the equation is of a fractional order, and has analytic solutions known (at least to us) only for $\nu=1$ and $\nu=2$. Although it is possible that a further insight can be gained by using the methods of fractional calculus \cite{FRACT}, we chose to solve the problem numerically, with the results presented in Fig.2.
For $\nu << 1$, the adiabatic state passes almost no time near threshold (cf. the inset in Fig.2), and $P^{stay}(\nu)$ tends to unity. 
As this time increases, we have $P^{stay}(\nu)\sim 1/\nu$, as prescribed by Eq.(\ref{g0}). The analysis is easily extended to asymmetric evolutions, as is illustrated in Fig. 6. 
\newline
With recent technological developments, it should be possible to verify our predictions in an experiment.
One straightforward choice would be the use of cold atoms in a laser induced quasi-one-dimensional trap \cite{Reiz}, which is manipulated as in "culling" or "squeezing", in order to bring one of its states to the continuum threshold. 
Another possibility is offered by studying the propagation of transverse modes in tapered wave guides \cite{WG1}-\cite{WG3}. Since narrowing of the guide lifts the energies of the quantised transverse motion, a massive particle or a photon, trapped in such a mode would have a similar chance of being lost to the continuum while passing the narrow region. A detailed analysis of wave guide propagation will be given in our forthcoming work.

\section{Acknowledgements}  Support of the Basque Government (Grant No. IT-472-10), and of the Ministerio de Economia y Competitividad of Spain (Grant No.FIS2015-67161-P) is gratefully acknowledged. 
%DS is also grateful to Gleb Gribakin and Gonzalo Muga for useful discussion of the subject. 

%%%%%%%%%%%%%%%%%%%%%%%%%%%%%%%%%%%%%%%%%%%%%%%%%%%%%%%%%%%%
\section{Appendix A: the energy distribution for a zero-range well }
Consider a particle of mass $\mu$ prepared in the bound state of a ZR well, evolving at a rate $v$ 
at $t=-T\to -\infty$. At $t\to \infty$, the deep impenetrable well divides the space at $x=0$, so that the continuum energy eigenstates are given by
\begin{equation}\label{ap1}
\phi_E(x,\mu)=(\mu/2\pi^2 E)^{1/4} \sin(\sqrt{2\mu E}|x|), 
\end{equation}
$\la\phi_E|\phi_{E'}\ra=\delta(E-E')$. The deep-lying bound state is decoupled from the continuum, and the ejected particles are described by the wave function $\psi_{cont}(x,t)=\int_0^{\infty}C(E)\phi(x,E)\exp(-iEt)dE$. For the (time independent) energy distribution we have 
\begin{equation}\label{ap2}
w(E|\mu,v)\equiv |\int \phi_E(x,\mu) \psi_{cont}(x,t)dx|^2.
\end{equation}
Consider next another particle of a mass $\mu'$, in a ZR well evolving at a different rate $v'$. We can also describe the  new system  by applying the transformations (\ref{b4}) to the old one. Thus, the ejected particles are described by $\psi'_{cont}(x,t)=\beta^{1/2} \psi_{cont}(\beta x, \alpha t)$. Inserting $\psi'_{cont}(x,t)$ into Eq.(\ref{ap2}) together with $\phi(x,E,\mu')$, and noting that $\mu'=\beta^2 \mu/\alpha$, we obtain 
 \begin{equation}\label{ap2}
w(E|\mu'',v'')=\alpha^{-1}w(E/\alpha|\mu',v'),
\end{equation}
where $\alpha(\mu',v''\mu,v)$ is  given by Eq.(\ref{b4}).
%%%%%%%%%%%%%%%%%%%%%%%%%%%%%%%%%%%%%%%%
\section{Appendix B} Let the trap be manipulated in such a way that the energy of the adiabatic bound state $E_n(t)$ changes with time according to
\begin{eqnarray}\label{bp1}
E_n(t)= -v^{2\nu}|t|^{2\nu}, \q for \q t<0, \n
 -v^{2\nu'}|t|^{2\nu'} \q for \q t>0.
\end{eqnarray}
Arguing as in Sect. II it is easy to show that the retention probability $P^{stay}(\nu,\nu')$ is given by 
  \begin{eqnarray}\label{bp2}
P_n^{stay}(v,\mu,\nu,\nu',W)=|\int \Psi(x,0,\nu)\Psi(x,0,\nu') dx|^2, \q\q
 \end{eqnarray}
 where $\Psi(x,t,\nu)$ is the result of evolving the initial state in such a manner that for $t<0$ its adiabatic energy changes according to  
 $E_n(t)= -v^{2\nu}|t|^{2\nu}$. As in Sect. VII, as $v\to 0$ we have $\Psi(x,t,\nu)\approx a_n(t,\nu,v)\phi_n(x,t,\nu)$ and 
 $\Psi(x,t,\nu')\approx a_n(t,\nu',v)\phi_n(x,t,\nu')$, where $\phi_n(x,t,\nu)$ denotes the corresponding adiabatic bound state.
 The coefficients $a_n(t,\nu,v)$ and $a_n(t,\nu',v)$ satisfy Eq.(\ref{q3}) with $k_n(vt,\nu)$ and $k_n(vt,\nu')$ respectively, 
 and in the limit $v\to0$ may depend only on $\nu$ and $\nu'$.
 Since $\phi_n(x,t,\nu)$ and $\phi_n(x,t,\nu')$ coincide at $t=0$, we should have $\int \phi_n(x,0,\nu)\phi_n(x,0,\nu')dx=1$, and 
 \begin{eqnarray}\label{bp3}
 \nonumber
P_n^{stay}(v\to 0,\mu,\nu,\nu',W)\to |a_n(t\to 0,\nu,v\to 0)|^2\\
\times |a_n(t\to 0,\nu',v\to 0)|^2,  \q
 \end{eqnarray}
Thus, also in the case of an asymmetric evolution the probability to remain in the bound state depends only on the powers $\nu$ and $\nu'$, and not on the particular shape of the potential, or the particle's mass.  Numerical examples are shown in Fig.6.
\begin{figure}
	\centering
		\includegraphics[width=9cm,height=6cm]{{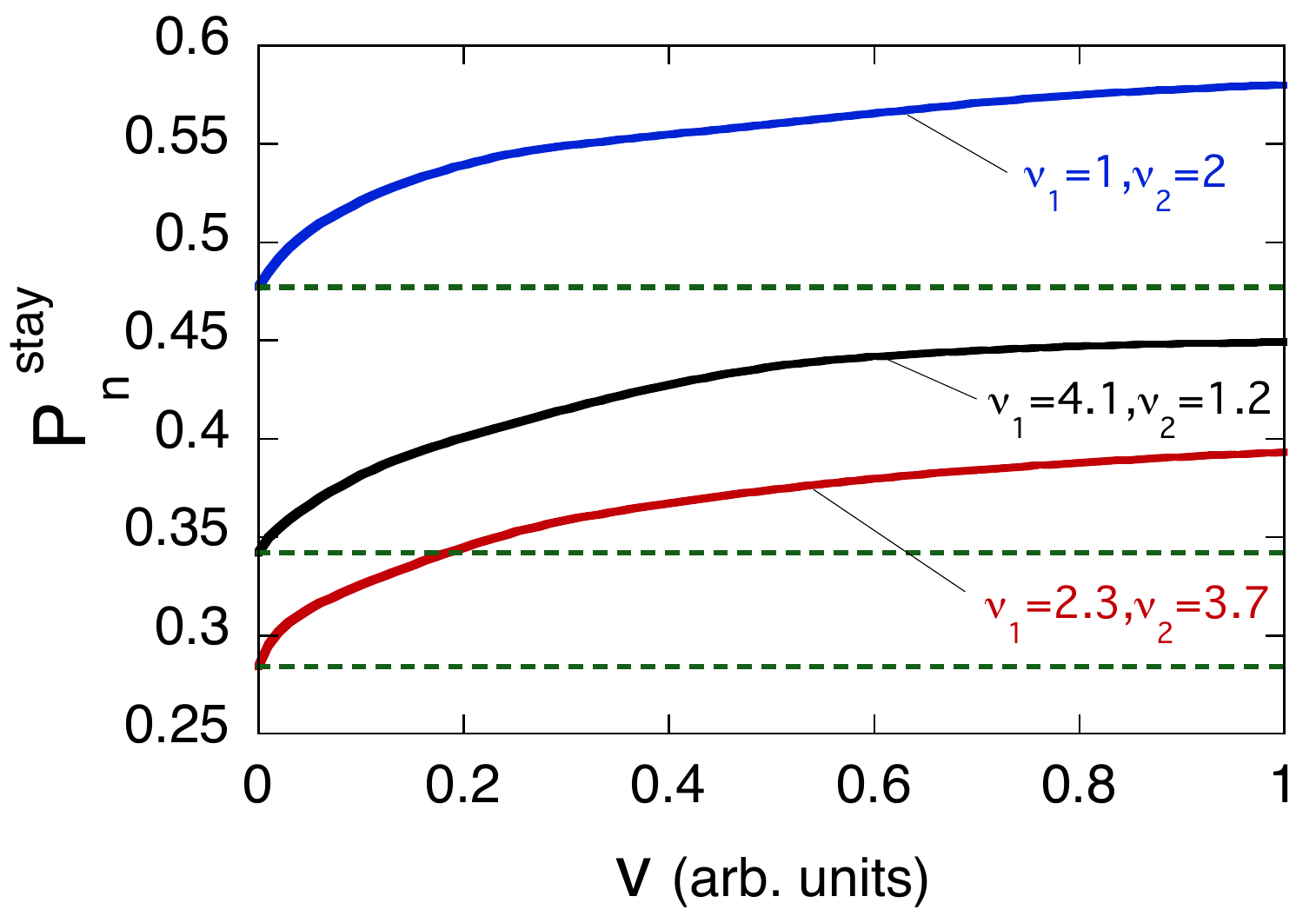}}
\caption{(Color online) The probability to remain in the ground  state, $n=0$, for $\mu=1$ and $W(x,t)=
[(vt)^{\nu_1} \theta(-t)+(vt)^{\nu_2} \theta(t)]W(x)$, $W(x)=W_I(x)$ (solid). Also shown by dashed lines are the corresponding results for the zero-range wells, $W(x)=\delta(x)$, independent of both $v$ and $\mu$.}
\label{fig:6}
\end{figure}

\end{document}